\documentclass[a4paper,11pt]{article}
\usepackage{pos}
\begin{document}
\title{POLAR-2, the next generation of GRB polarization detector}
\author*[a]{Nicolas Produit}
\author[b]{Merlin Kole}
\author[b]{Xin wu}
\author[b]{Nicolas De Angelis}
\author[a]{Hancheng Li}
\author[c]{Dominik Rybka}
\author[c]{Agnieszka Pollo}
\author[c]{Slawomir Mianowski}
\author[d]{Jochen Greiner}
\author[d]{J. Michael Burgess}
\author[e]{Jianchao Sun}
\author[e]{Shuang-Nan Zhang}
\affiliation[a]{Department of Astronomy, University of Geneva, \\Chemin d’Ecogia 16, CH-1290 Versoix, Switzerland}
\affiliation[b]{DPNC, University of Geneva, quai Ernest-Ansermet 24, 1205 Geneva, Switzerland}
\affiliation[c]{National Centre for Nuclear Research ul. A. Soltana 7, 05-400 Otwock, Swierk, Poland}
\affiliation[d]{Max-Planck-Institut fur extraterrestrische Physik, Giessenbachstrasse 1, D-85748 Garching, Germany}
\affiliation[e]{Key Laboratory of Particle Astrophysics, Institute of High Energy Physics, Chinese Academy of
Sciences, Beijing, China, 100049}
\emailAdd{Nicolas.Produit@unige.ch}
\abstract{The POLAR-2 Gamma-Ray Burst (GRB) Polarimetry mission is a follow-up to the successful POLAR mission. POLAR collected six months of data in 2016-2017 on board the Tiangong-2 Chinese Space laboratory. From a polarization study on 14 GRBs, POLAR measured an overall low polarization and a hint for an unexpected complexity in the time evolution of polarization during GRBs. Energy-dependent measurements of the GRB polarization will be presented by N. de Angelis in GA21-09 (August 2nd). These results demonstrate the need for measurements with significantly improved accuracy. Moreover, the recent discovery of gravitational waves and their connection to GRBs justifies a high-precision GRB polarimeter that can provide both high-precision polarimetry and detection of very faint GRBs. The POLAR-2 polarimeter is based on the same Compton scattering measurement principle as POLAR, but with an extended energy range and an order of magnitude increase in total effective area for polarized events. Proposed and developed by a joint effort of Switzerland, China, Poland and Germany, the device was selected for installation on the China Space Station and is scheduled to start operation for at least 2 years in 2025.}
\ConferenceLogo{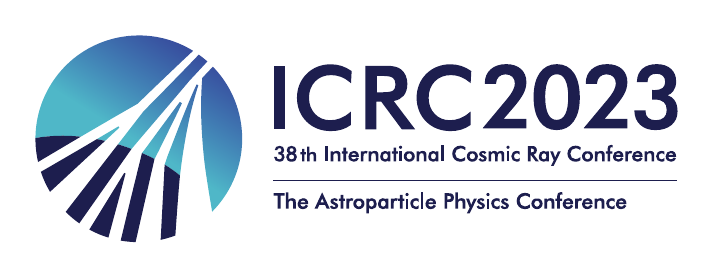}

\FullConference{%
38th International Cosmic Ray Conference (ICRC2023)\\
  26 July - 3 August, 2023\\
  Nagoya, Japan}

\maketitle

\section{Introduction}
\begin{figure}[!htbp]
	\begin{center}
	\includegraphics[width=\textwidth]{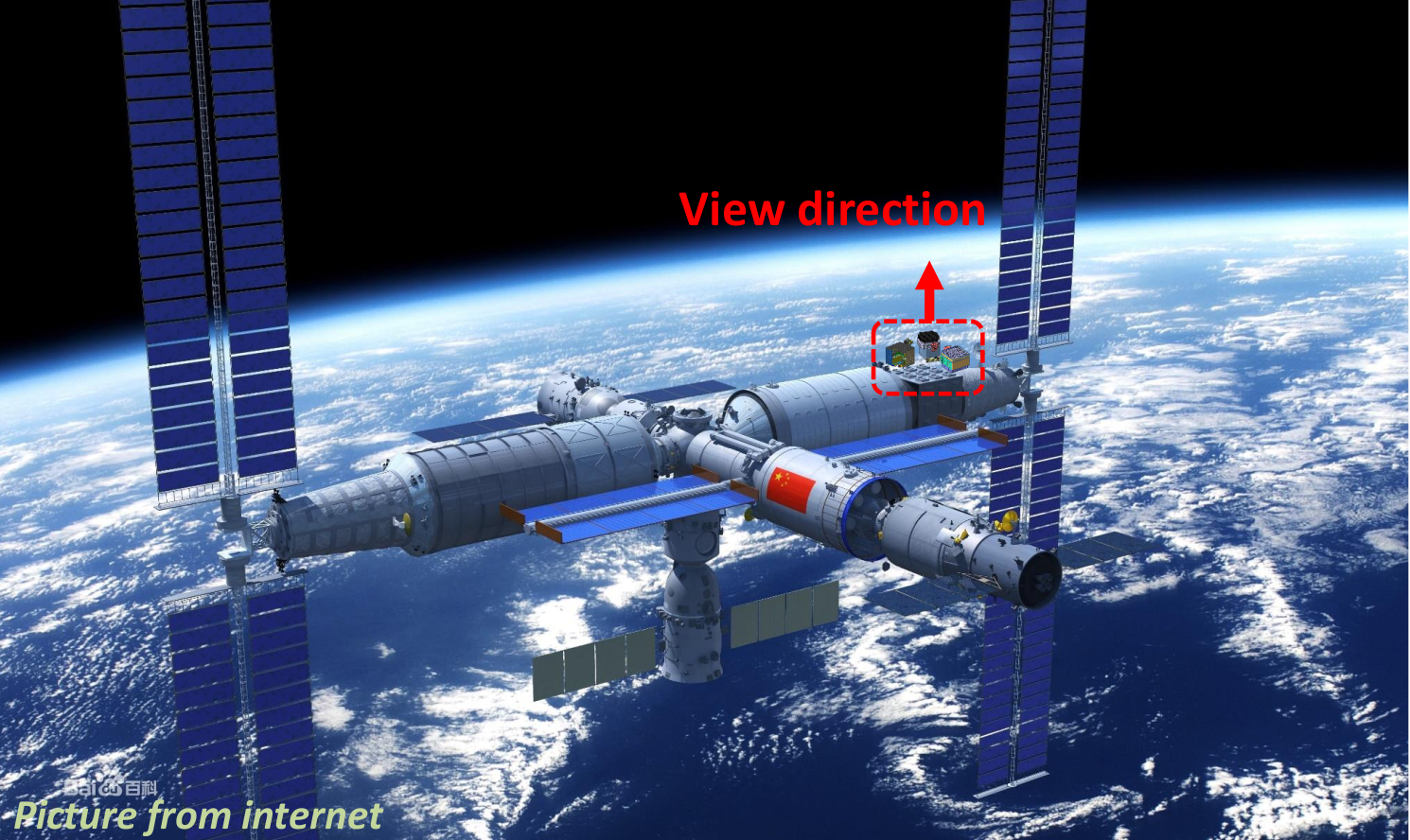}
		\caption{Completed in space in 2023, the Tiangong Chinese space station is shown along with the allocated position of the high energy polarimeter, and the possible positions of the imager and the low energy polarimeter. The 3 payloads will continuously point away from Earth, thereby observing approximately half the sky continuously (with exception from passages through the South Atlantic Anomaly).}
		\label{fig:fith}
	\end{center}
\end{figure}

POLAR-2  on the China Space Station (CSS) will consist of 3 sub instruments: the high energy polarimeter, which is described in this paper, but also an imager and a low energy polarimeter. The former is fully approved for launch and construction of the flight model will commence later this year, the two other payloads remain in the design phase (see fig. \ref{fig:fith}). The high energy polarimeter construction takes place in Switzerland and Poland, while the imager and low energy polarimeter, which uses wide field of view detectors similar to those on eXTP and IXPE, will be produced in China.

\section{Measurement Principle}

\begin{figure}[h]
	\begin{center}
    \includegraphics[height=.4\textwidth]{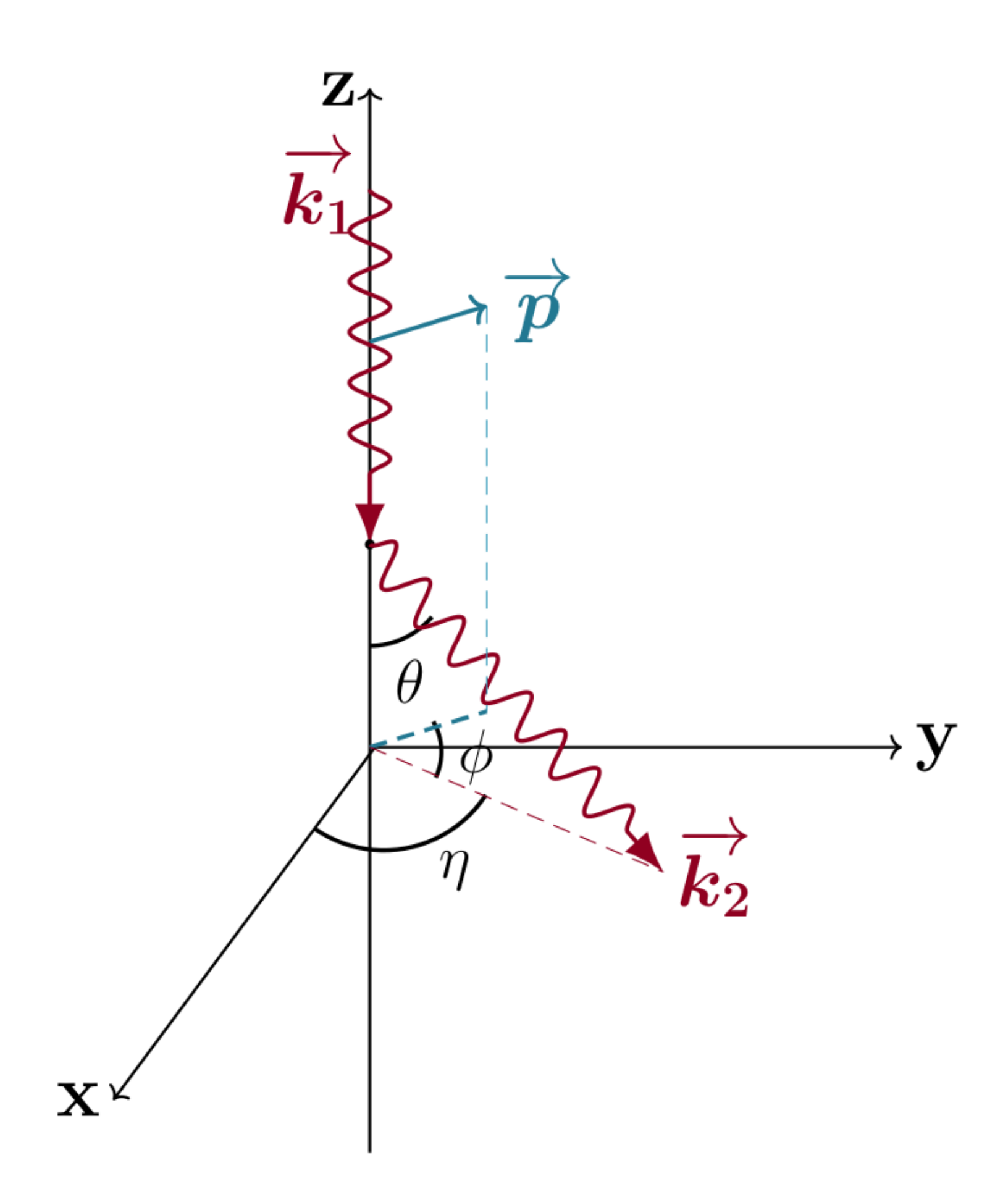}\hspace*{0.5cm}\includegraphics[height=.4\textwidth]{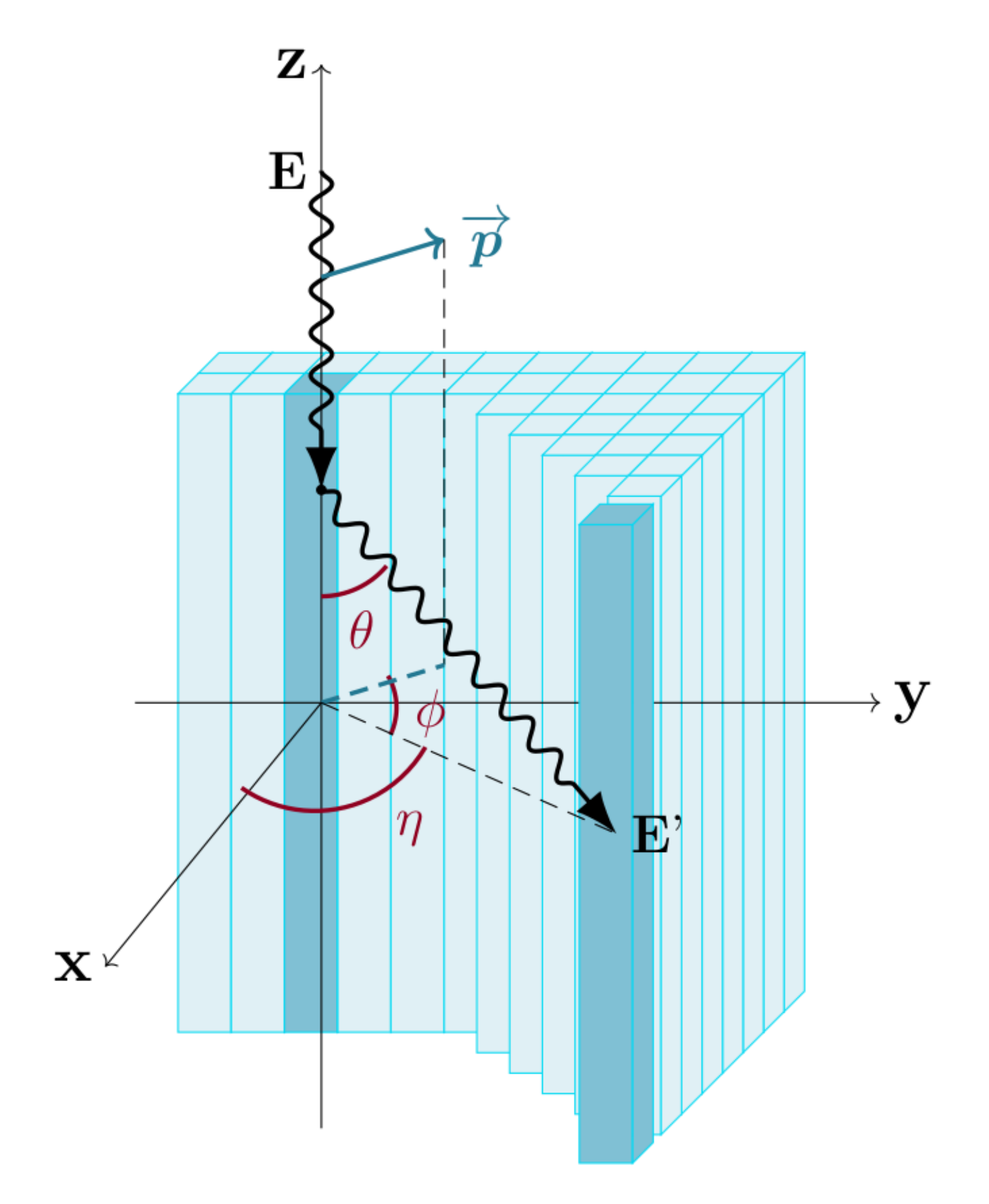}\hspace*{0.5cm}\includegraphics[height=.4\textwidth]{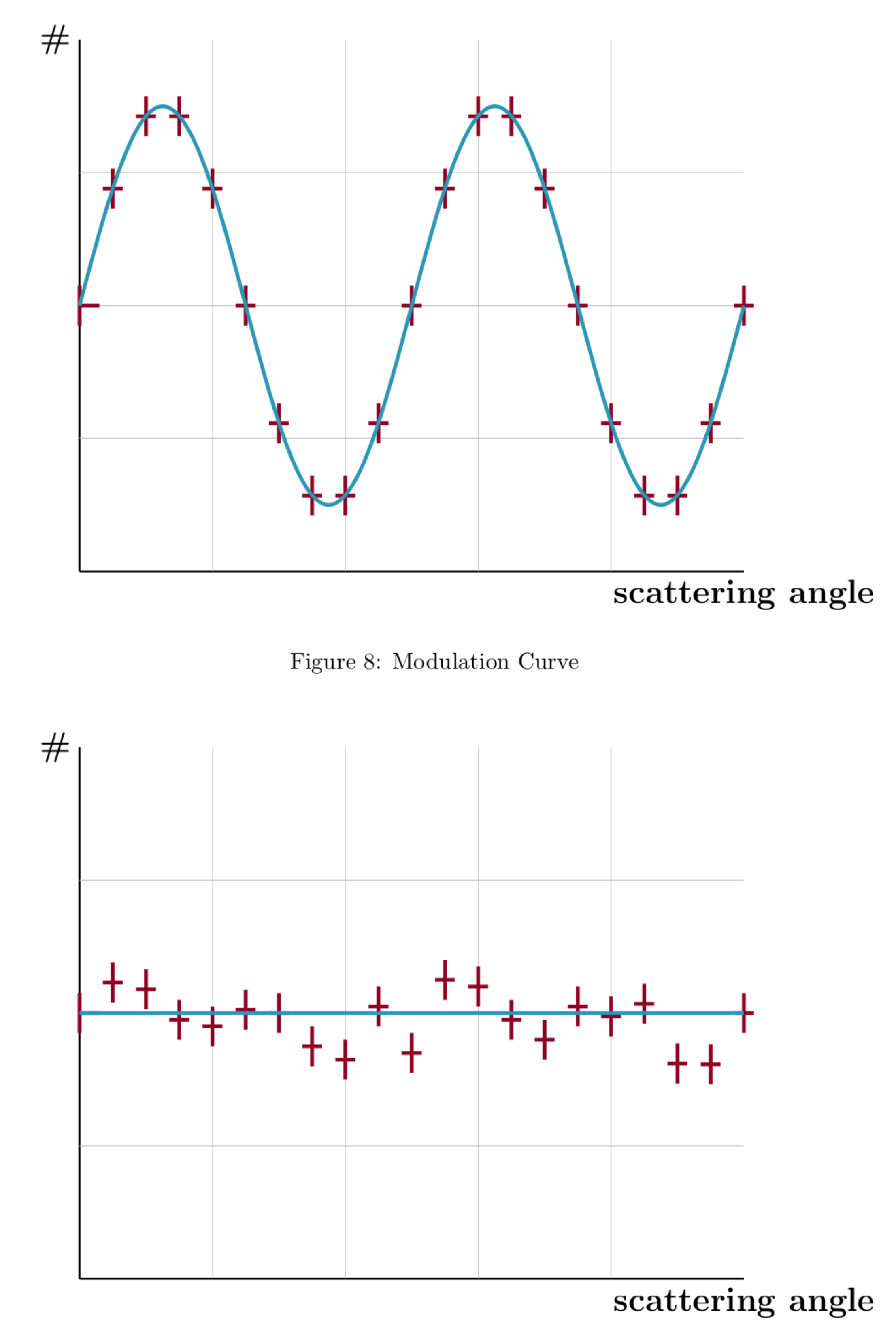}
		\caption{Left: An illustration of a Compton scattering interaction with the incoming and outgoing vector of the photon as $\vec{k}$ and its polarization vector $\vec{p}$ along with the two scattering angles $\theta$ and $\phi$. Middle: the idea of measuring the $\phi$ angle using a segmented detector array. Right: Scattering angle distributions, or modulation curves, for a polarized (top) and unpolarized (bottom) flux. }
		\label{fig:CE}
	\end{center}
\end{figure}

POLAR-2 will measure the polarization of the incoming GRB flux by exploiting the dependence of the azimuthal Compton scattering angle on their polarization vector. This dependency can be seen in the Klein-Nishina equation of the cross-section for Compton scattering:

\begin{equation} 
\mathrm{\frac{d\sigma}{d\Omega}=\frac{r_o^2}{2}\frac{E'^2}{E^2}\left(\frac{E'}{E}+\frac{E}{E'}-2\sin^2\theta \cos^2\phi\right).}
\end{equation}

In here, E is the energy of the incoming photon, $E'$ that of the outgoing photon, $r_o$ is the classical radius of the electron, $\theta$ is the forward scattering angle as illustrated also on the left of figure \ref{fig:CE} and $\phi$ is the azimuthal Compton scattering angle. The $\cos^2\phi$ term in this equation causes the photons to preferentially scatter perpendicular to their polarization vector. 

POLAR-2 exploits this by using a segmented detector array in which the photons can scatter in one detector element before interacting in a second. By knowing the relative position of the two detector elements the Compton scattering angle can be calculated. By doing this for a flux of incoming photons and producing a histogram of these angles (as shown on the right in figure \ref{fig:CE}, a so called modulation curve can be produced. In case of a polarized incoming flux the modulation curve will show an azimuthal dependency with a period of $180^\circ$ degrees, while if the flux is inpolarized each photon will scatter in a random direction, thereby making this curve flat. For the polarized case the polarization degree of the incoming flux can be derived from the amplitude of the modulation, whereas the phase is correlated to the polarization angle. It should be noted that in reality this analysis is significantly more complex and much was learned on the various issues and pitfalls of polarization analysis with the POLAR data. An overview on the POLAR analysis method and the many encoutered issues can be found in \cite{POLAR_analysis}.

\section{Design}

\subsection{Detector Modules}

\begin{figure}[h]
	\begin{center}
    \includegraphics[height=.4\textwidth]{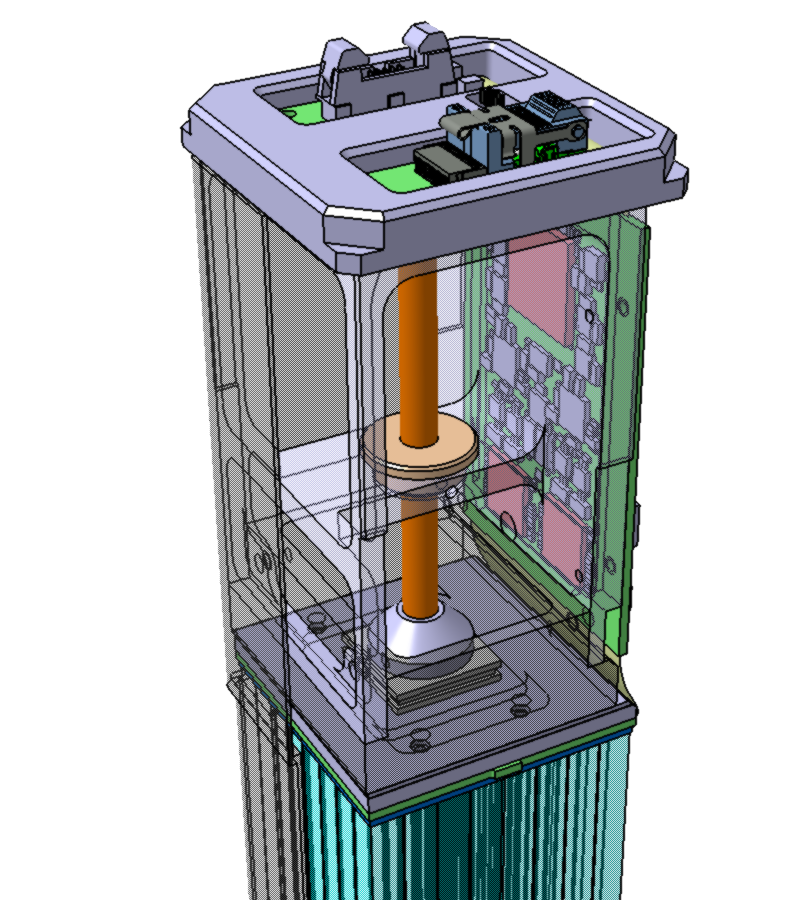}\hspace*{0.5cm}\includegraphics[height=.4\textwidth]{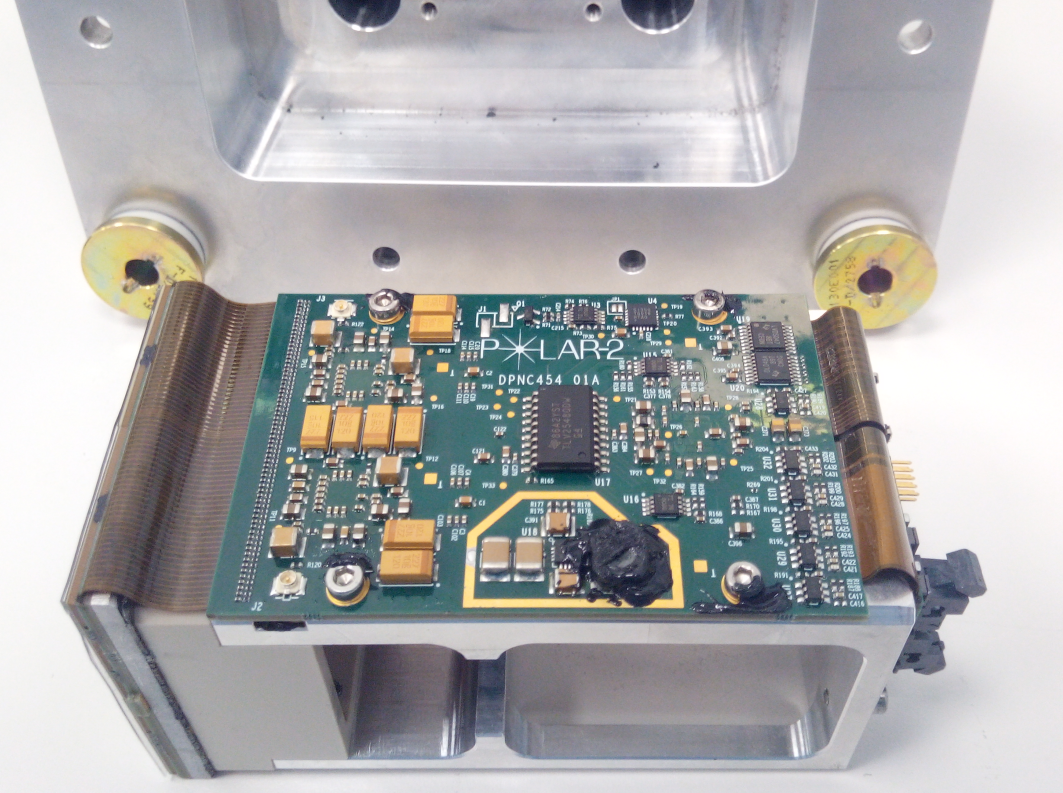}
		\caption{Left: The mechanical design of a POLAR-2 DM. The scintillators are shown inside of the carbon fiber exterior mechanics along with the scintillators connected to the FEE. Next to the FEE is a copper pipe which leads the heat from a Peltier element away from the SiPMs towards the POLAR-2 radiators. Right: A POLAR-2 FEE prototype.}
		\label{fig:FEE}
	\end{center}
\end{figure}

The POLAR-2 detector consists of 6400 plastic scintillator bars with dimensions of $5.9\times5.9\times125\,\mathrm{mm^3}$.  The readout of these bars is changed from using Multi-Anode PhotoMultiplier Tubes (MAPMTs) in POLAR to Silicon Photomultipliers (SiPMs in POLAR-2). The scintillator bars are combined in groups of 64 to form a total of 100 independent detector modules (DMs). The DMs contain the 64 scintillators, as well as 4 $4\times4$ arrays of SiPMs (S13361-6075NE-04 MPPCs by Hamamatsu) as well as the front-end electronics (FEE) responsible for the readout and digitization of the SiPMs and the combination with the back-end electronics. A full POLAR-2 FEE prototype is shown along with the CAD design which shows the scintillators inside of the carbon fiber housing in figure \ref{fig:FEE}.

\subsection{Scintillators}

The plastic scintillators consist of EJ-248M by Scionix, this is the same scintillator material as that used in POLAR. Although for POLAR this choice was based on the larger thermal resistance of the material compared to materials like EJ-200 which have a high light yield, for POLAR-2 this thermal resistance is not a requirement. Instead, measurements with prototypes indicated that the EJ-248M shows a better performance for the number of optical photons reaching the SiPM compared to that with EJ-200. Although the intrinsic light yield for EJ-248M is $9.2$ optical photons per keV compared to 10.0 for EJ-200, the surface quality of EJ-248M is significantly better. This, based on disucssions with the manufacturer, is a result of the material being harder than EJ-200, thereby allowing for smoother polishing and therefore better reflection inside of the scintillator. As a result of these tests, which will be discussed in detail in an upcoming paper, EJ-248M was selected for POLAR-2.

Each scintillator is wrapped in 2 layers of highly reflective material. First each scintillator is fully wrapped on the 4 sides in Vikuiti by 3M. This ensures a significantly larger number of photons reaching the bottom of the scintillator. To avoid any optical cross talk a layer of Claryl foil (an aluminiumized plastic) is placed outside of the vikuiti in between the scintillators to stop the approximately $1\%$ of the photons which manage to penetrate the Vikuiti. The bars are finally alligned using a 3d printed plastic grid to provide an alignment within $50\mu m$ with the SiPM arrays. A good optical coupling between the scintillators and the SiPMs is ensured using a silicon pad of $200\mu m$ thickness which is grown, in house using a method developed at the University of Geneva for POLAR-2, on top of the SiPMs.

Through studies of the measured POLAR background in space, an optimization of the scintillator length was also performed. By shortening the bars from 178 mm as they were in POLAR to 125 mm, the effective area is slightly reduced. However, the background rate is significantly more reduced, thereby decreasing the Minimal Detectable Polarization (MDP).

\subsection{Front-End Electronics}

The readout and operation of the of the SiPMs is performed using the FEE. The FEE contains 2 Citiroc-1A ASICs (each with 32 readuot channels) which handle the amplification and readout of the SiPMs. The trigger logic and communication with the ASICs is handled using a simple FPGA (IGLOO by Microsemi). This FPGA furthermore controls the HV bias to the SiPMs, which is automatically corrected for changes in temperature. Finally, the FPGA processes the data by performing zero-surpression and packaging of the data before sending it to the back-end electronics which handles the communication between the 100 FEEs which make up POLAR-2. 

Finally, the FEE contain a Peltier element which is placed on the back of the SiPM array to cool these down. The Peltier element is foreseen to consume between 0.3 and 1.5 W depending on the outcome of future thermal simulations. The Peltier element is connected to a copper pipe which leads the heat it creates away from the SiPMs towards the instruments radiators. 

The power consumption of one FEE, without the Peltier element, is 1.7 W.

\subsection{Overall Design Changes compared to POLAR}

\begin{figure}[h]
	\begin{center}
    \includegraphics[height=.4\textwidth]{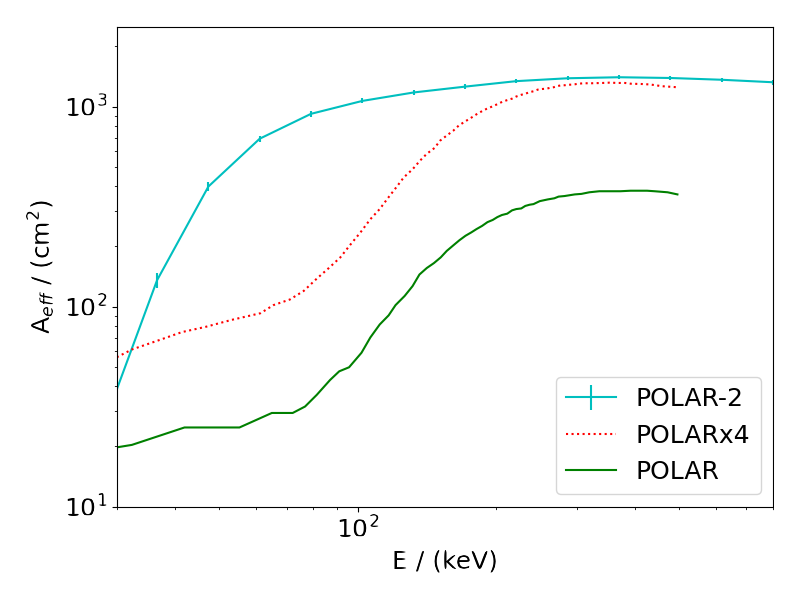}\hspace*{0.5cm}\includegraphics[height=.4\textwidth]{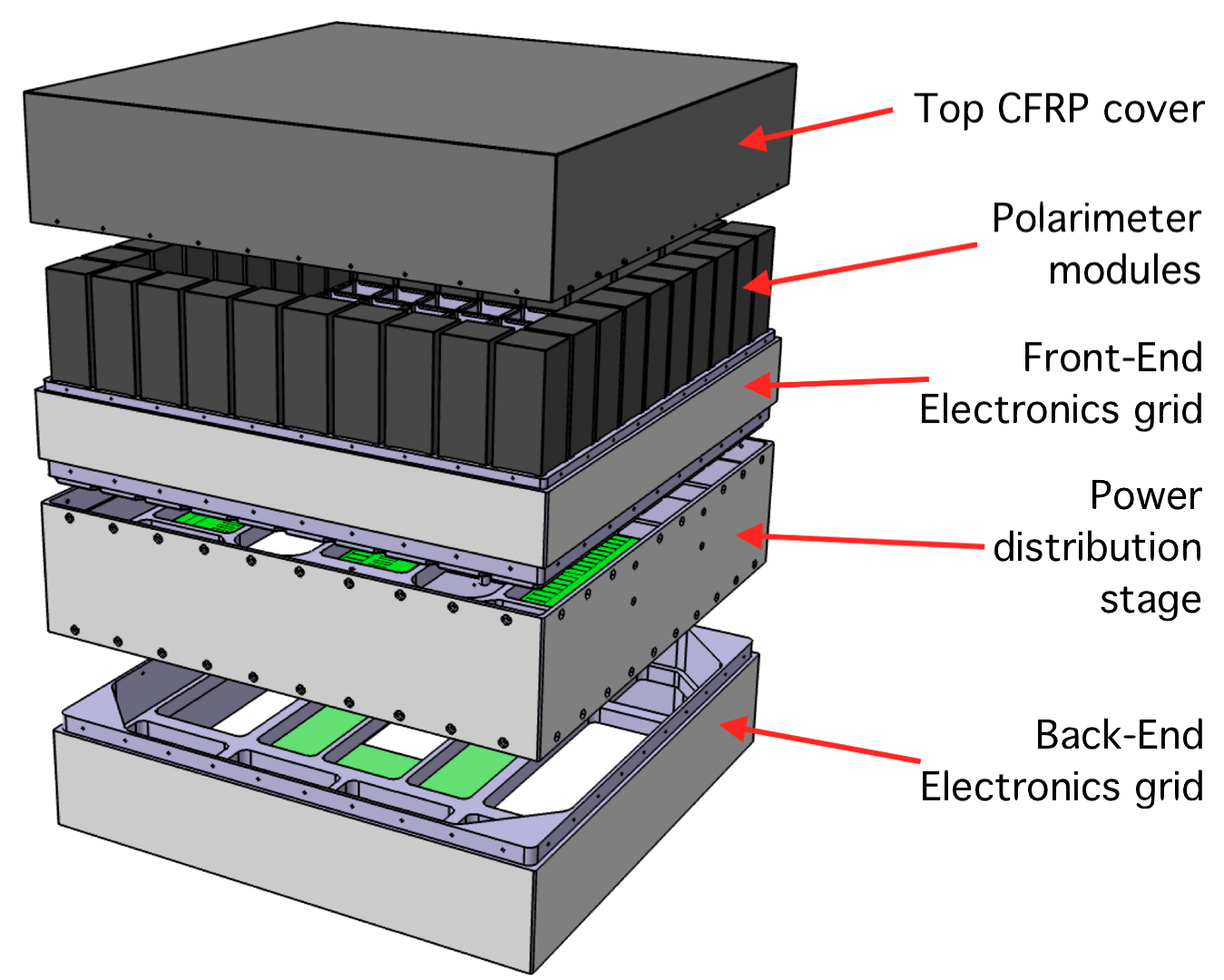}
		\caption{Left: the effective area versus initial photon energy. Green shows the effective area of POLAR, and red that of POLAR when simply increasing its size by a factor of 4. Blue shows the POLAR-2 effective area. It can be seen that POLAR-2 is significantly more sensitive at low energies thanks to the SiPM technology, Right: an exploded view of the POLAR-2 high energy polarimeter. We can see the 100 modules in black and the electronics in green}
		\label{fig:mech}
	\end{center}
\end{figure}

POLAR-2's design is an improvement over that of POLAR in several ways. Firstly, the size is increased by a factor of 4, resulting in a total of 100 detector modules (25 for POLAR) each containing 64 scintillating bars. The second large improvement, is the use of SiPMs instead of MPPC. This change along with an optimization of the scintillator bar shape, increases the light yield from 0.3 photo-electrons (p.e.)/keV in POLAR, to 1.6 p.e./keV for POLAR-2. Due to the dark noise of SiPMs the trigger threshold for a single scintillator readout channel has to be approximately 4 p.e. to ensure an instrument level trigger rate below $100\,\mathrm{Hz}$. Combining this with the light yield we can see that the low energy threshold for POLAR-2 is around $2.5\,\mathrm{keV}$ whereas it was $\approx12.5\,\mathrm{keV}$ for POLAR. This change thereby significantly improves the sensitivity of POLAR-2 at low energies. This is illustrated on the left of figure \ref{fig:mech} where the effective area of POLAR, as well as a simply version of POLAR scaled up by a factor 4, is compared to that of POLAR-2. It can be seen that especially at lower energies the increase in effective area is significant.

The increase in the energy range is accomplished thanks to the ASIC used in POLAR-2. The CITIROC has two independent amplifiers for each channel which, in the case of POLAR-2, are both readout and digitized for each trigger. By setting the gain of the two amplifiers at different levels one can be used for detailed measurements at low energies while the second is used to extend the dynamic range up to 800 keV (compared to approximately 350 keV for POLAR). It should be noted that most photons will deposit energy in 2 different channels, thereby making the dynamic range of a single channel not equal to the full energy range of the detector.

A downside of the use of SiPMs is their dark noise which requires one to operate the instrument at low temperature (our goal is -20$^\circ$C, it was around +20$^\circ$C for POLAR). For this purpose, all the design is optimized towards low temperatures, for example, the FEE design is optimized to reduce heat from the electronics to move towards the SiPM. In addition all the external surfaces of POLAR-2 are painted with high reflectivity and high emissivity paint which was specifically tested and space qualified for POLAR-2. The full POLAR-2 design is shown on the right of figure \ref{fig:mech} which shows the 100 detector modules at the top (covered by a carbon fibre cover which serves as a passive shield for low energy electrons and for mechanical protection). Below the 100 DMs a low voltage power supply which provides power to the DMs and the BEE is placed on an aluminium grid. Finally at the bottom the BEE is placed in its own aluminium grid which will be mounted on the space station using a robotic arm. The aluminium outside will be painted such that it serves as a radiator.

Finally, SiPMs have been shown to degrade in space due to radiation damage \cite{DeAngelis2023}. Based on simulations we know that the low energy threshold will increase due to this by approximately 1 p.e. ($\approx 0.6\,\mathrm{keV}$ per year. In order to mitigate this, we have also tested that the irradiation damage can be recuperated by bringing the SiPMs to a high temperature for a couple of hours. Heating to 60$^\circ$C using simple resistors is implemented and will be performed once per year for about 1 day in orbit
\cite{DeAngelis2023}.

\section{Performance}

The first prototype detector modules of POLAR-2 have been tested in the lab using radioactive sources as well as during a dedicated calibration campaign at the ESRF synchrotron in April 2023 \cite{Kole2023}. This data, which was taken using 3 prototype detector modules, was used to firstly measure the light yield performance of each scintillator readout channel. The light yield for the optimum configuration was found to be 1.6 p.e./keV and therefore matches that expected from previous simulations. In addition, it was found that whereas for POLAR the use of MAPMTs results in significant optical crosstalk (around $10-15\%$ between channels, for POLAR-2 design the optical cross talk is insignificant as it well below $1\%$. This significantly increases POLAR-2's sensitivity to polarization and simplifies data analysis. Finally, it was found that at room temperature a single channel threshold (which forces a trigger of the full instrument in case one channel in a module goes above this) of $8\,\mathrm{keV}$ could be achieved, while for the double channel threshold (which triggers the readout if 2 channels in a single module exceed this) this can be set to $5\,\mathrm{keV}$. In orbit, where the temperature is expected to be $0\,C^\circ$ this can be lowered to 5 and 3 keV based on the current performance.

\begin{figure}[h]
	\begin{center}
    \includegraphics[height=.3\textwidth]{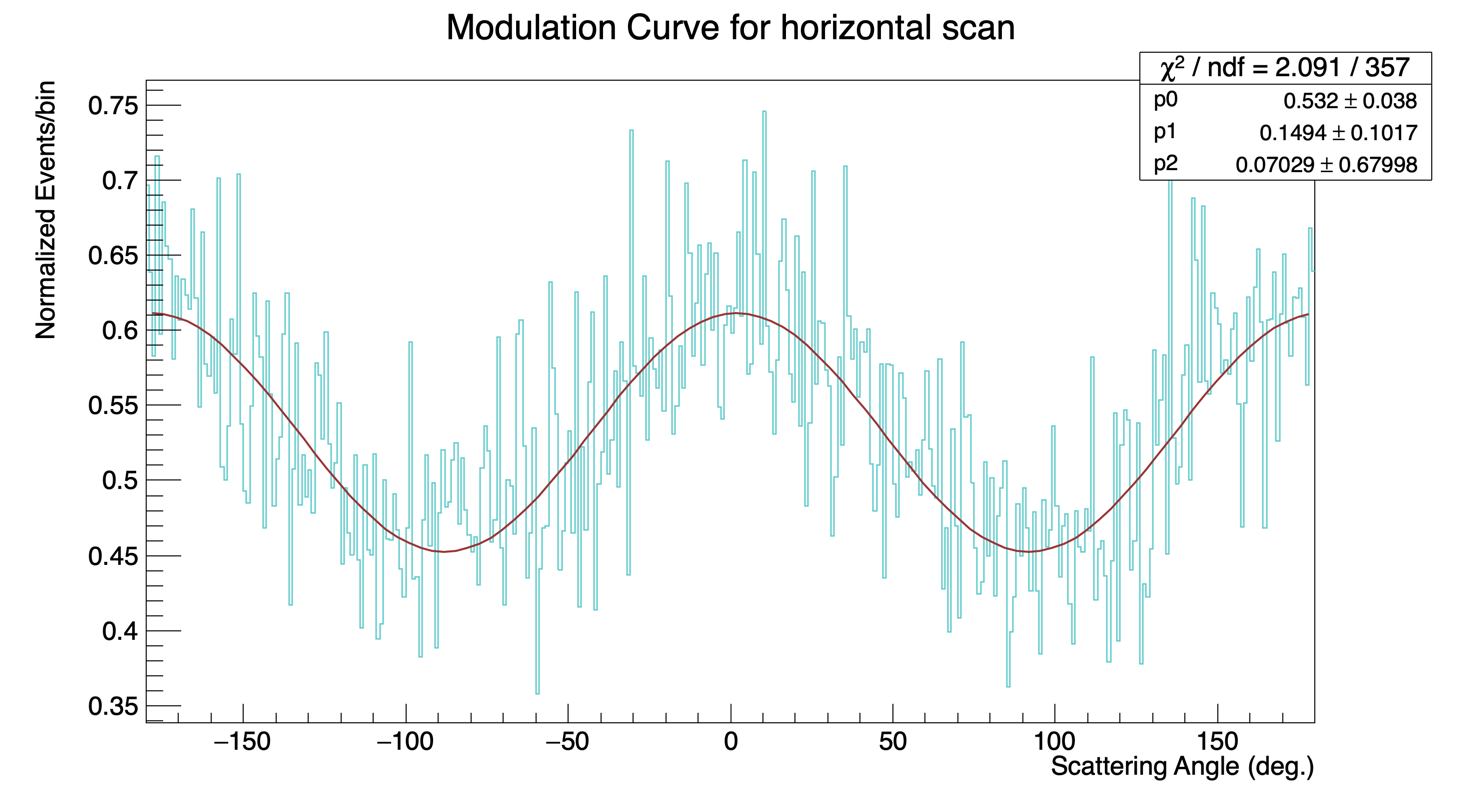}\hspace*{0.5cm}\includegraphics[height=.3\textwidth]{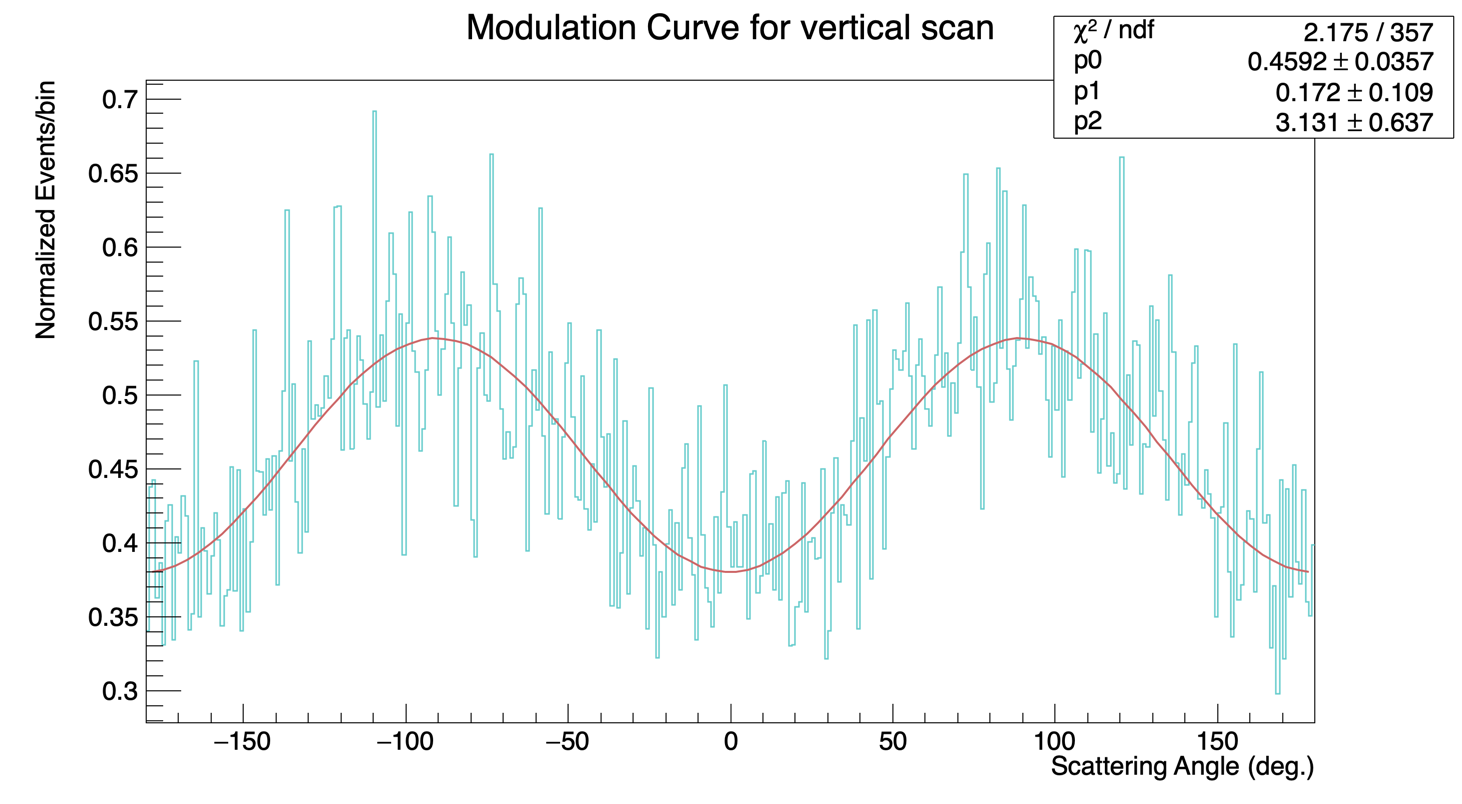}
		\caption{Left: The corrected modulation curve for a single detector module for a 40 keV beam as measured at ESRF. Right: The same as on the left but with the incoming beam polarized perpendicular. The 2 first fit parameters correspond to the mean and the $M_{100}$ (as a fraction).}
		\label{fig:mod_40keV}
	\end{center}
\end{figure}

In addition to these studies the data from ESRF is currently being used to study the performance of the detector to polarized emission. In total calibration data was used to measure the instrument response at 40, 60, 80, 100 and 120 keV for different incoming and polarization angles. The results, which are still under study, indicate a good agreement with the simulated instrument responses. Preliminary results of the response of a single detector module to a 40 keV polarized beam (in two different directions) can be seen in figure \ref{fig:mod_40keV}. While more optimization of the data analysis needs to be performed, these results already show a significant improvement over POLAR which was unable to measure polarization at such low energies. These first results show a modulation factor of $\approx15\%$ at $40\,\mathrm{keV}$ for a single module. For 60, 100 and 120 keV, the current results show modulation factors of $20\%, $30\% and $27\%$ respecitvally. Simulations using GEANT4 \cite{G4} will be performed over the coming weeks to extrapolate how this modulation factor increases when moving from 1 to 100 detector modules. \\

\begin{figure}[t]
	\begin{center}
    \includegraphics[height=.4\textwidth]{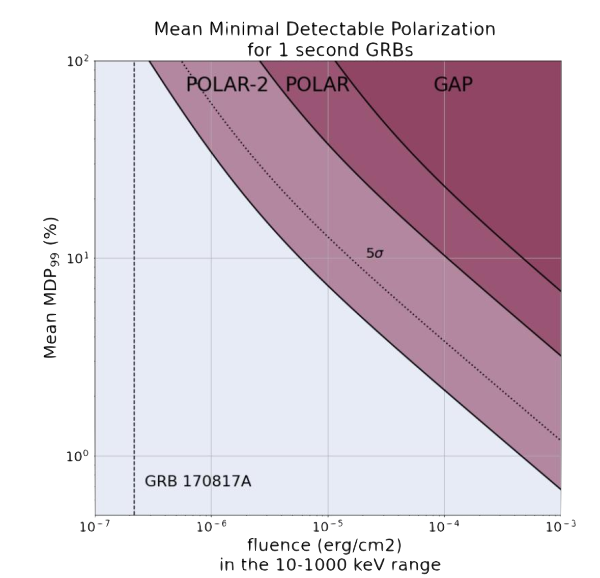}\hspace*{0.5cm}\includegraphics[height=.4\textwidth]{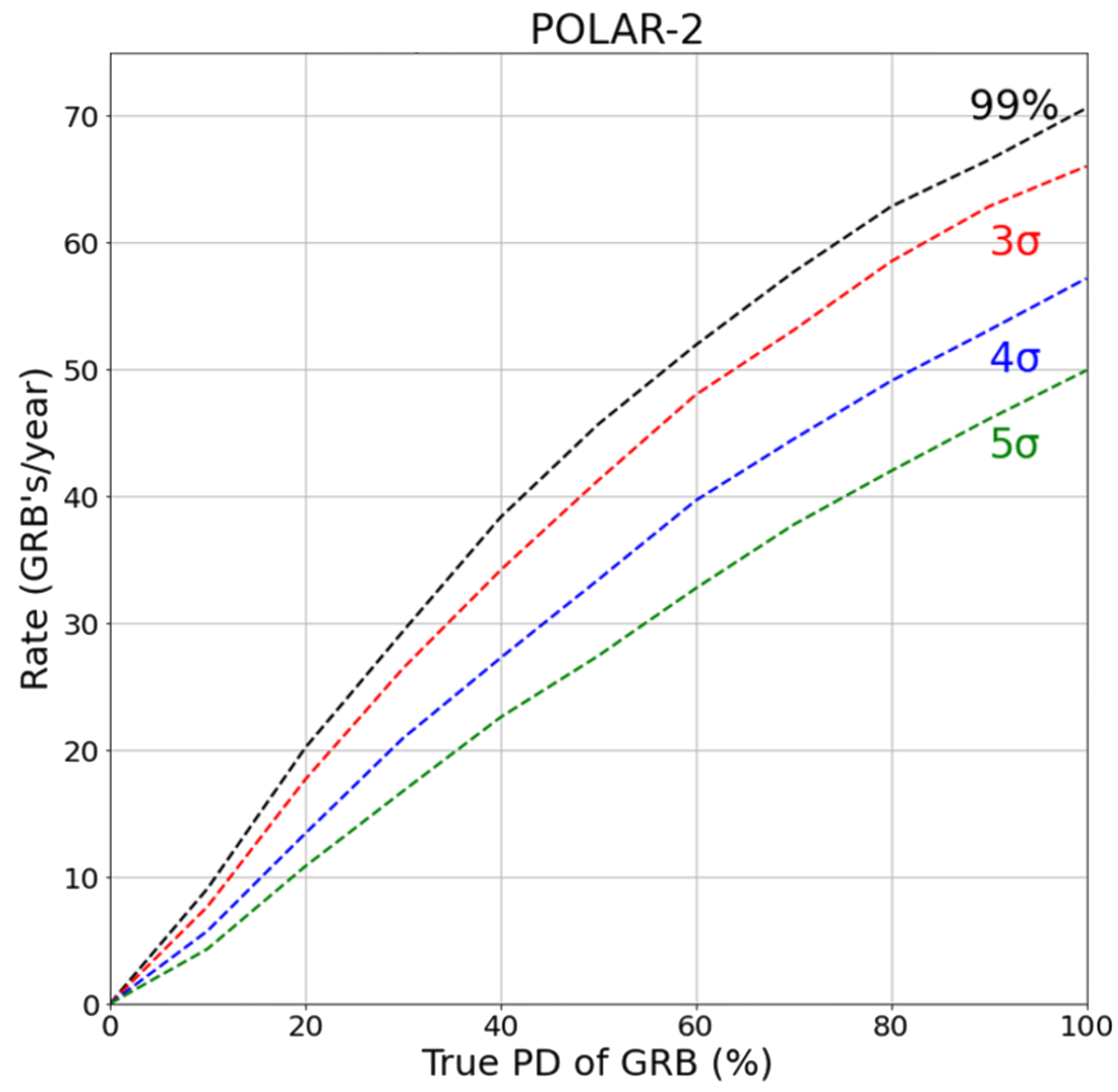}
		\caption{Left: The Minimal Detectable Polarization (in log-scale) as a function of the GRB brightness, for 1 seconds GRBs, for POLAR-2, POLAR and GAP. The brightness of the extremely weak GRB 170817A is shown as a reference. Taken from \cite{GW_paper}. Right: The number of GRBs for which a significant polarization measurement can be performed per year as a function of their intrinsic polarization. The significance of the measurement is shown for various significance levels. Taken from \cite{Gill2021}.}
		\label{fig:sense}
	\end{center}
\end{figure}

\bigskip
The POLAR-2 Monte-Carlo simulations are now being updated using the calibration data from ESRF. The preliminary simulations were already used to predict the scientific performance of POLAR-2. The Minimal Detectable Polarization (the lowest level of intrinsic polarization which can be distinguished from being unpolarized) is shown as a function of the GRB fluence for POLAR-2 on the left in figure \ref{fig:sense}). The MDP is compared here with the sensitivity of its predecessor POLAR and that of GAP (the first dedicated GRB polarimeter launched in 2010). When combining this with existing GRB catalogs and known orbital elements of the spacecraft, one can predict the number of GRBs (see right side of fig. \ref{fig:sense}) for which a signicicant polarization detection can be performed, as a function of the instrinsic GRB polarization, per year. 

Apart from polarimetry, POLAR-2 will be able to perform real-time analysis of the incoming angle of GRB, and therefore its location on the sky, as well as their spectrum through access to a GPU on the space station (see HAGRID poster and paper [PGA1-06] in this conference). As uncertainties on the spectrum and location induce systematic errors, this allows POLAR-2 to completely independently measure the linear polarization for all observed GRBs. In addition, we are investigating the possibility of POLAR-2 to send alerts with localization and spectral info to the ground within 2 minutes of the onset of the GRB. The fast alert system, combined with the very large effective area will allow POLAR-2 to play an important role in multi-messenger astrophysics. 

\bibliographystyle{JHEP}
\bibliography{my-bib-database}
\end{document}